\documentclass[referee]{risaV2}

\usepackage[figuresright]{rotating}
\usepackage{multirow}

\begin{document}


\title[Topological Performance Measures as Surrogates for Physical Flow Models]{Topological Performance Measures as Surrogates for Physical Flow Models for Risk and Vulnerability Analysis for Electric Power Systems}

\author[LaRocca, Johansson, Hassel, and Guikema]{
Sarah LaRocca,$^{1}$\affiliation{Department of Geography and Environmental Engineering, Johns Hopkins University, Baltimore, MD, USA}
Jonas Johansson,$^{2,3}$\affiliation{Lund University Centre for Risk Assessment and Management (LUCRAM), Lund, Sweden}\affiliation{Department of Measurement Technology and Industrial Electrical Engineering, Lund University, Lund, Sweden}
Henrik Hassel,$^{2,4}$\affiliation{Department of Fire Safety Engineering and Systems Safety, Lund University, Lund, Sweden}
Seth Guikema$^{1}$ 
} 


\begin{abstract}

\large{\noindent{\textbf{Abstract}}}

Critical infrastructure systems must be both robust and resilient in order to ensure the functioning of society. To improve the performance of such systems, we often use risk and vulnerability analysis to find and address system weaknesses. A critical component of such analyses is the ability to accurately determine the negative consequences of various types of failures in the system. Numerous mathematical and simulation models exist which can be used to this end. However, there are relatively few studies comparing the implications of using different modeling approaches in the context of comprehensive risk analysis of critical infrastructures. Thus in this paper, we suggest a classification of these models, which span from simple topologically-oriented models to advanced physical flow-based models. Here, we focus on electric power systems and present a study aimed at understanding the tradeoffs between simplicity and fidelity in models used in the context of risk analysis. Specifically, the purpose of this 
paper is to compare performances measures achieved with a spectrum of approaches typically used for risk and vulnerability analysis of electric power systems and evaluate if more simplified topological measures can be combined using statistical methods to be used as a surrogate for physical flow models. The results of our work provide guidance as to appropriate models or combination of models to use when analyzing large-scale critical infrastructure systems, where simulation times quickly become insurmountable when using more advanced models, severely limiting the extent of analyses that can be performed.

\end{abstract}




\maketitle


\section{Introduction}
\label{sec:intro}

Critical infrastructure systems form the foundation for the economic prosperity, security, and public health of the modern world \cite{Rinaldi2004}. As such, failures within these complex systems can pose a significant threat to society. Unfortunately, failures in infrastructure systems occur relatively frequently, arising from a variety of sources including natural disasters, terrorism, and accidents. Such failures, which may at first seem isolated or insignificant, have the potential for far-reaching and devastating consequences. Thus, understanding the reliability and vulnerability of such systems has become an increasingly significant concern of decision-makers in both the public and private realms. Understanding system vulnerabilities and reliability is of particular concern as utilities have become increasingly interested in proactive risk management after recent events such as Hurricanes Irene and Sandy in the U.S. and major winter storms that have had significant impacts on power systems.

A crucial factor in conducting useful reliability and vulnerability analyses is the ability to accurately characterize the consequences of failures within the system.  Understanding a system's robustness - that is, the degree of sensitivity of system performance to failures - allows us to identify and address critical weaknesses in the system.  This understanding is generally gained through the use of a system performance model, and the fidelity of these models vary significantly. For example, for electric power infrastructure, performance models vary from purely topological-based models that do not incorporate the engineering or physical aspects of the system performance to complex AC power flow models based on the physical and engineering details of the system. If we use models which incorrectly predict system performance, our assessments are likely to give rise to sub-optimal management decisions for the infrastructure system in question.  Unfortunately, the accuracy of such models is often taken for 
granted when assessing the robustness (i.e., the opposite of vulnerabilty) of infrastructure systems.  Thus, in this work, our goal is to understand the implications of using models of varying complexity for evaluating infrastructure system performance.  To limit the scope of our work, we will focus specifically on electric power systems.
 
The following approach is commonly used for assessing the robustness of infrastructure systems: 1) modeling the initial performance of the infrastructure system of interest; 2) simulating various types of failures in these systems; and 3) evaluating the consequences of the failures by use of some measure of system performance \cite{Watts1998, Motter2002b, Crucitti2004a, Chassin2005, Holmgren2006a, Wang2009a}.  However, for a given infrastructure system, there are numerous mathematical and simulation models which can be used to this end; in this paper, we refer to such models as \emph{functional models} \cite{Johansson2012}.  Additionally, system robustness can be quantified by a variety of \emph{performance measures}.

Functional models currently in use for electric power system analysis range in complexity from pure topological approaches to physics-based models of AC power flows. Strict topological models only use information about the network structure (\emph{i.e.}, nodes and edges) to describe the behavior of the system, ignoring physical constraints such as the physics governing power flow.  This means that some important factors affecting system performance are neglected \cite{Grubesic2008}; in return, the models are computationally efficient, meaning that it is possible to analyze large systems and a large number of contingencies within feasible computational times. Another benefit is that very little information about the system is needed to perform risk and vulnerability studies.  Additionally, topological models require significantly less data about the system than physics-based models.  Such physics-based models, often used by power engineers, incorporate capacity limits of system components as well as the 
physics governing power flow (\emph{i.e.}, Kirchoff's laws).  These models will provide the most accurate representation of a power system, however, their computational complexity often makes their use impractical, particularly when modeling large systems and analyzing many failure scenarios.

There has been little research aimed at systematically evaluating the impact of using different functional models for assessing electric power system robustness. Hines et al. \citet{Hines2010} compare different models for evaluating electric power systems. They conclude that topological models may lead to misleading results as compared to performance estimates from a DC-linearized load flow model. However, they did not compare their results to those of a full AC power flow model and they only considered two topological performance measures. Overbye et al. \citet{Overbye2004} compare the use of DC-linearized and AC power flow models for setting Locational Marginal Prices (LMP), concluding that the two models produce satisfactorily similar results. However, it is difficult to generalize their findings to the present context since the study was not conducted with regard to analyzing robustness. In addition, they did not look into simpler topological models, and they only addressed failures scenarios involving a 
single system component which are not likely to provide a full picture of system robustness. Finally, Chen et al. \citet{Chen2010} suggest a hybrid approach for modeling cascading failures that includes a DC-linearized power flow model. However, they only provide a comparison to a single topological performance measure (efficiency) and the comparison made is not as systematic as is necessary to enable a clear conclusion to be drawn.

In this paper, we present a study that aims to improve our understanding of the tradeoffs between simplicity and fidelity of these models in the context of assessing infrastructure system robustness. More specifically, the goal of the paper is to compare different functional models used to estimate the performance of electric power systems in order to evaluate how well simplified functional models are able to capture the behavior of the systems when exposed to perturbations.  Finally, we aim to develop a method which combines the strengths of existing approaches to yield a model that accurately reflects system behavior while still maintaining computational feasibility.


\section{Classification of functional models}
\label{sec:models}

As mentioned in the introduction, the functional models used in existing studies of infrastructure robustness span from  very simple to very advanced. In this section we propose a general classification of such approaches, consisting of four classes of increasingly advanced functional models: topological models with undifferentiated components; topological models with differentiated components; simplistic capacity models; and physical flow models. We describe each of these classes in the following subsections, focusing on approaches used to assess electric power system robustness. It should be noted, however, that a similar classification can be used for other types of technical infrastructure models.


\subsection{Topological models, undifferentiated components}
\label{ssec:models:undiftopo}

Many existing studies of infrastructure robustness employ topological functional models based on network theory.  Such models are a particularly valuable tool for assessing infrastructure robustness, because most infrastructure systems naturally take the form of a network.  Topological models disregard physical flows in the system, instead representing the system abstractly as a collection of nodes and edges.  In the simplest category of topological models, there is no differentiation between components in the system; that is, different functions within the set of nodes or edges are ignored \cite{Holme2002, Motter2002b, Latora2005, Holmgren2006a, Rosas-Casals2007, Winkler2010}.  When modeling power systems, this means that no distinction is made between buses, substations, or generators - all are treated simply as nodes (overhead power lines and underground cables are treated simply as edges).


\subsection{Topological models, differentiated components}
\label{ssec:models:diftopo}

Neglecting to differentiate between types of system components may provide an inaccurate representation of reality, particularly if the components are actually highly heterogeneous (e.g., have significantly different functions). Therefore, a second, more complex, category of topological models is often used, incorporating details about the various functions of the system components.  For power systems, a commonly used approach is to model the system as a network consisting of three types of nodes: generators, stations, and load points; another approach is to simply differentiate between in-feed and load nodes \cite{Albert2004, Holmgren2006a, Duenas-Osorio2009, Winkler2010}.


\subsection{Simplistic capacity models}
\label{ssec:models:capacity}

Simplistic capacity models are a class of functional models which use network flow methods combined with actual system data to represent loads and capacities in the system.  Because such methods do not attempt to incorporate \emph{physical} flow modeling (e.g., hydraulic modeling or power flow analysis), but instead rely on a network-based approach, these models should still be seen as predominantly topological. Several simplistic capacity models have been used for analyzing power system robustness. Wang et al. \citet{Wang2011} develop a functional model which incorporates information about maximum load and generator capacity in the system along with line impedances with a traditional topological approach, resulting in a concept they call `electrical betweenness.'  Another approach, presented in J\"{o}nsson et al. \citet{Jonsson2008} uses capacity values for all in-feed nodes (i.e., generators), as well as demand at load nodes (i.e., distribution substations) to calculate the amount of power not supplied to 
substations.  This functional model relies a network search algorithm to `push' capacity of an in-feed node through the network to load nodes, rather than conducting a complete load flow analysis in accordance with Kirchoff's laws.


\subsection{Physical flow models}
\label{ssec:models:physflow}

The topological approaches described above may not fully capture the details regarding the physical flow in the systems under study. However, the fundamental physical laws governing the flows in different types of infrastructure are typically well-known, and are therefore easy to include in a functional model, at least conceptually.  Modeling such physical flows does come at a cost, though; both computational times and initial data requirements are likely to increase when using such a functional model. For electric power systems, physical flows are typically addressed by the use of a DC-linearized or AC load flow model to evaluate the steady-state conditions of the system. Several previous studies have, to varying extent, incorporated DC or AC load flow analysis in assessing infrastructure robustness and reliability \cite{Dobson2001, Carreras2002a, Dobson2002, Dobson2003, Song2005, Pepyne2007, Arianos2009, Chen2010}.
 

\subsection{Performance measures}
\label{ssec:models:performance}

For any given functional model, there may be multiple measures that can be used to quantify system performance.  For example, when using a topological model with undifferentiated components (i.e. representing the system as a network with no additional information except the relationships between nodes and edges), a variety of network theoretic measures can be selected to describe system performance, including size of the largest connected subgraph, average path length,  and network diameter.  Or, when using a physical flow model, such as DC load flow for an electric power system, performance could be quantified as unsupplied load or the number of customers without power.  When comparing functional models, it is important to also consider the corresponding performance measure being used.  Thus, in this work, we evaluate functional model-performance measure \emph{pairs}.


%


\section{Methods}
\label{sec:methods}


\subsection{Test system}
\label{ssec:methods:testsys}

In this work, we use the one-area IEEE Reliability Test System-1996 (RTS96), a bulk power transmission system (230 and 138 kV) including generation, transmission, and loads \cite{Grigg1999}. As a test system designed specifically for reliability studies, the description of RTS96 includes detailed data on generation reliability and capacity, transmission system reliability and capacity, and load curves with respect to both yearly and daily variation \cite{Grigg1999}. The system consists of 24 buses (nodes) and 38 branches (edges). The annualized peak power demand is 2850 MW in total; annual and daily fluctuations of loads are not taken into account here. Aggregated generation capacity is 3405 MW. We use the 24-hour emergency power rating of lines for line capacity.


\subsection{Functional models and performance measures}
\label{ssec:methods:models}

As discussed previously, there are a number of functional models and performance measures which can be used to analyze the robustness of infrastructure systems.  In this work, we test 9 different functional model-performance measure pairs using the IEEE RTS-96 system described above, as summarized in Table~\ref{tab:funcmodel}.  Although many of these functional models and performance measures are flexible enough to incorporating the potential for cascading failures, here we focus only on `static' versions of these models. The following sections describe in detail the functional models and corresponding performance measures used in our analysis.


\begin{table*}
\centering
\def\~{\hphantom{0}}
\caption{Functional models and performance measures used in analysis.}
\label{tab:funcmodel}
\begin{tabular*}{\textwidth}{@{}l@{\extracolsep{\fill}}l@{\extracolsep{\fill}}l@{\extracolsep{\fill}}}
\Hline
\textbf{Functional model} & \textbf{Performance measure} & \textbf{Label}\\
\hline
\multirow{3}{*}{Topological, undifferentiated components} & Largest connected component & LCSG\\
 & Diameter & D \\
 & Efficiency & E \\
\hline
\multirow{4}{*}{Topological, differentiated components} & Efficiency, pairs of in-feed and load nodes & EN\\
 & Efficiency, pairs of in-feed and load nodes, weighted by impedance & ENE \\
 & Connectivity loss & CL \\
 & Power connection loss & PCL \\
\hline
Simplistic capacity & Power not supplied & PNS \\
\hline
\multirow{2}{*}{Physical flow } & Power not supplied, based on DC power flow & DC \\
 & Power not supplied, based on AC power flow & AC \\
\Hline
\end{tabular*}
\end{table*}


\subsubsection{Topological models, undifferentiated components}
\label{ssec:methods:models:undiftopo}

In existing studies of electric power system robustness using a topological model with undifferentiated components, a variety of network theory-based performance measures have been suggested \cite{Albert2004, Holmgren2006a, Rosas-Casals2007, Sole2008, Arianos2009, Winkler2010, Hines2011}.  Here, we evaluate three of these performance measures: largest connected subgraph; network diameter; and network efficiency.  These performance measures, as used in conjunction with a topological function model with undifferentiated components, are described below.

\paragraph{\normalfont\textbf{Largest connected subgraph (LCSG)}}
The largest connected subgraph in a graph is defined as the largest subgraph in which a path exists between all pairs of nodes.  Then, the size of the largest connected subgraph is defined as:
\begin{equation}
 S_{LCSG} = N_{LCSG},
\end{equation}
where $N_{LCSG}$ is the number of nodes in the largest subgraph.

\paragraph{\normalfont\textbf{Diameter (D)}}
The diameter of a network is defined as the `longest shortest path' in the network, that is:
\begin{equation}
 D = \textrm{max}_{i,j} d_{ij},
\end{equation}
where $d_{ij}$ is the length of the shortest path (i.e., number of edges) between node $i$ and node $j$.
 
\paragraph{\normalfont\textbf{Efficiency (E)}}
Network efficiency, also known as average inverse path length, is defined as follows:
\begin{equation}
 E = \frac{1}{N(N-1)}\sum_{i,j}\frac{1}{d_{ij}},
\end{equation}
where $N$ is the number of nodes in the network and $d_{ij}$ is the length of the shortest path between node $i$ and node $j$.
 

\subsubsection{Topological models, differentiated components}
\label{ssec:methods:models:diftopo}

As previously mentioned, not differentiating between different types of system components may result in a misrepresentation of true system behavior. In order to overcome this limitation, several topologically-based performance measures have been used in existing studies to account for the fact that all nodes and edges do not have the same function \cite{Albert2004, Johansson2007}. Additionally, we propose two new topological measures that we hypothesize might more accurately capture the performance of electric power systems.  These performance measures, both existing and newly proposed, are described below.

\paragraph{\normalfont\textbf{Efficiency, pairs of in-feed and load nodes (EN)}}
As described above, network efficiency is calculated based on the shortest paths between all pairs of nodes in the network. However, in an electric power system it may not be particularly relevant whether pairs of load-nodes are well connected unless they are also well-connected to those nodes that inject the electric flow into the system (e.g. generators and transformers). Thus, our first newly proposed measure of network efficiency is calculated as with the traditional measure of network efficiency, $E$, described above, with the exception that only paths between in-feed and load nodes are considered.  That is,  
\begin{equation}
 EN = \frac{1}{N(N-1)}\sum_{i \in N_{F},j \in N_{L}}\frac{1}{d_{ij}},
\end{equation}
 where $N$ is the total number of nodes in the network, $N_F$ is the set of in-feed nodes, $N_L$ is the set of load nodes, and $d_{ij}$ is the length of the shortest path between node $i$ and node $j$.
 
\paragraph{\normalfont\textbf{Efficiency, pairs of in-feed and load nodes, weighted by impedance (ENE)}}
Here, we suggest a second new measure incorporating `electrical distance,' that is, line impedance, into the shortest path calculations. Our second measure of network efficiency is calculated as $EN$, with the addition that path length is weighted by electrical line impedance.  So, we have:
\begin{equation}
 ENE = \frac{1}{N(N-1)}\sum_{i \in N_{F},j \in N_{L}}\frac{1}{d_{ij}|Z_{ij}|},
\end{equation}
where $N$ is the total number of nodes in the network, $N_F$ is the set of in-feed nodes, $N_L$ is the set of load nodes, $d_{ij}$ is the length of the shortest path between node $i$ and node $j$, and $|Z_{ij}|$ is the magnitude of the impedance of path $ij$.
 
\paragraph{\normalfont\textbf{Connectivity loss (CL)}}
Connectivity loss is a topologically-based performance measure for electric power systems that was first proposed in Albert et al. \citet{Albert2004}.  It describes the `ability of distribution substations to receive power from the generators,' and is defined as follows:
\begin{equation}
 CL = 1 - \frac{1}{N_D}\sum_{i}^{N_D}\frac{N_{G}^i}{N_G},
\end{equation}
where $N_G$ is the total number of generators, $N_D$ is the total number of distribution substations, and $N_G^i$ is the number of generators connected to substation $i$.
 
\paragraph{\normalfont\textbf{Power connection loss (PCL)}}
Power connection loss was first described by Johansson et al. \citet{Johansson2007} as the aggregate load at nodes that do not have any connection to an in-feed node, such as a generator or transformer. It is thus defined as:
\begin{equation}
 PCL = \sum_{i \in NC} \textrm{load}_i, 
\end{equation}
where $NC$ is the set of nodes that do not have any connection to an in-feed node and load$_i$ is the load at node $i$.


\subsubsection{Simplistic capacity models}
\label{ssec:methods:models:capacity}

We evaluate a simplistic capacity model for electric power systems that was first presented in J\"{o}nsson et al. \citet{Jonsson2008}. This network flows-based algorithm, which is used to calculate total amount of real power not supplied to substations without incorporating Kirchoff's laws, is described below.

\paragraph{\normalfont\textbf{Power not supplied (PNS)}}
This method requires capacity values for all in-feed nodes (i.e., generators), as well as demand at load nodes (i.e., distribution substations). Power not supplied is calculated as follows: 1) select initial in-feed node; 2) “push” capacity of in-feed node through network using a breadth-first search; 3) subtract substation loads from initial capacity of in-feed node when a substation is reached and flag substation as supplied; 4) continue distributing capacity of initial in-feed node until it has been consumed; 5) select another in-feed node; 6) return to step 1, repeating until all connected substations are supplied or all available in-feed capacity is consumed; 7) power not supplied is equal to the total substation load that is not supplied.  Thus, we have:
\begin{equation}
 PNS = \sum_i^n \textrm{load}_{demanded_i} - \textrm{load}_{supplied_i},
\end{equation}
where $n$ is the number of nodes in the network, $\textrm{load}_{demanded_i}$ is the demand at node $i$ and $\textrm{load}_{supplied_i}$ is the load supplied to node $i$.


\subsubsection{Physical flow models}
\label{ssec:methods:models:physflow}

For electric power systems, physical flow-based functional models involve load flow analysis to evaluate the steady-state conditions of the system, either using a DC-linearized approximation or a full AC power flow model. The most accurate way to represent the physical flow of power in an electric power system is to use an AC load flow model. However, AC power flow is described by nonlinear equations for which convergent solutions are often difficult to obtain; solving AC power flow requires significant computational resources and time which are often prohibitive, particularly in large-scale simulations. As a result, a DC-linearized approximation, which only considers the flow of real power, ignoring reactive power, is often used to approximate AC power flow. The relative simplicity of the DC equations combined with their linearity allows a direct (i.e. non-iterative) solution to be obtained quickly.  

\paragraph{\normalfont\textbf{Power not supplied, based on DC load flow analysis (DC)}, and \textbf{power not supplied, based on AC load flow analysis (AC)
}}

To perform both DC and AC load flow modeling, we use a Matlab package called Matpower \cite{Zimmerman2011}, which was developed through the Power Systems Engineering Research Center (PSERC). Matpower allows for calculation of DC linearized power flow, AC power flow, DC linearized optimal power flow (DC OPF), and AC optimal power flow (AC OPF). Optimal power flow is determined through an objective function which minimizes generation and unsupplied load costs and includes constraints such as branch capacity and voltage limits. Here we use the optimal power flow algorithm for both the AC and DC models, curtailing load until a solution can be attained. If a solution cannot be found which satisfies the constraints, all load in the system or subsystem (if the initial system has split into several subsystems) \cite{IEEE1995, Billinton1996, Arnborg1997, Arini1999, Ladhani2004}. We measure system performance as the total amount of load (active power only) curtailed as a result of failures in the system.
 \\
The generation, loading, and branch-limits used were provided with the test system. The settings for busbar voltage limits were 1.1 p.u. for the upper limit and 0.7 p.u. for the lower limit.  This relatively low value for the lower voltage limit was selected because in this work we are calculating load flow for a severely strained system. However, a system operating at below 0.7 p.u. is likely to experience a voltage collapse, in accordance with Taylor \cite{Taylor1994}. The loads in the system were designated as negative generators and associated with a large negative cost (piecewise linear cost function with the settings $x_0=0$, $y_0=0$, $x_1=-P_{load}$, and $y_1=-10000P_{load}$). The generation cost was set with low positive values (polynomial cost function with nominal values for $c_2=1$, $c_1=1$, and $c_0=0$) \footnote{In the rare occasion of convergence problems with the optimization algorithm, different $c_2$ values were selected in attempts to find a converging solution, where $c_2 \in \{0.001, 0.01,
 0.1, 2, 4, 5, 18, 32, 64\}$}.
 

\subsection{Failure scenarios}
\label{ssec:methods:failures}

Since our goal in this work is to evaluate the effectiveness of various functional models in the context of assessing infrastructure system vulnerability, we develop a set of failure scenarios, or strains, which our network experiences. Most topological studies of power system vulnerability focus on node removals, so we assess node failures here. However, in real power systems, overload-related failures are more likely to occur in lines than in buses, and thus it is important to also address edge failures. We simulate each type of failure independently; that is, in one set of scenarios we consider node failures and in another we consider edge failures.

In order to limit the scope of our work, we only evaluate scenarios in which system components fail randomly.  To generate a given random failure scenario, we use a uniform random number generator to sequentially select nodes or edges for removal from operation, resulting in a strain vector, or failure scenario vector, containing a random ordering of all nodes or edges in the the system. We repeat this process 1,000 times for both nodes and edges, resulting in two strain matrices (one for nodes and one for edges) consisting of 1,000 vectors of randomly ordered component failures. We then use the strain matrices to simulate failures in our test system and evaluate the subsequent system performance using the functional model-performance measure pairs described above in Section~\ref{ssec:methods:models}.


\subsection{Statistical analysis}
\label{ssec:methods:stats}

Ideally, the reference for comparing the results from using different functional models and performance measures should be empirical results from the system of interest. However, since we are conducting our analysis on a fictitious test system, such data do not exist \footnote{Even when analyzing a real system, it is highly unlikely that one would be able to obtain empirical data for more than a few failure scenarios.}.  Therefore, we assume that the most advanced functional model (\emph{i.e.,} the full AC load flow model) corresponds most closely with the true performance of the system. For the AC load flow functional model, our performance measure is the load curtailed (real power) in the system as a fraction of the initial load in the system, that is, the percent change in load that the system is able to meet after a given failure scenario. Because different performance measures are used for other functional models (\emph{e.g.,} network diameter for a pure topological approach) and do not directly 
correspond to load curtailed, we standardize all other performance measures to the range [0,1] in order to carry out our comparisons.

For each of the nine functional model-performance measure pairs described above, we fit simple linear regression models with load curtailed as based on AC load flow analysis as the response variable.  Table~\ref{tab:simplinreg} summarizes the models for each functional model-performance measure. We also fit multiple linear regression models using six different combinations of functional model-performance measure pairs as covariates. After fitting each of these initial models, we iteratively remove all covariates from the model that are not statistically significant. That is, for a given model, we remove the explanatory variable with the highest p-value, refit the model, and repeat until all variables are statistically significant at the level of $\alpha = 0.05$.  Table~\ref{tab:multlinreg} presents each combination of covariates used to develop our multiple linear regression models.  As with the simple linear regression models (Table~\ref{tab:simplinreg}), we use six different sets of data to fit six 
independent multiple linear regression models for each of the covariate combinations in Table ~\ref{tab:multlinreg}.
 

\begin{table}
\centering
\def\~{\hphantom{0}}
\caption{Summary of simple linear regression models developed for each functional model-performance measure pair.}
\label{tab:simplinreg}
\begin{tabular*}{\columnwidth}{@{}c@{\extracolsep{\fill}}c@{\extracolsep{\fill}}c@{\extracolsep{\fill}}}
\Hline
{\textbf{Model}} & {\textbf{Element removed}} & {\textbf{Number removed}}\\
\hline
a & Nodes & 1 \\
b & Nodes & 3 \\
c & Nodes & 5 \\
d & Edges & 5 \\
e & Edges & 7 \\
f & Edges & 9 \\
\Hline
\end{tabular*}
\end{table}


\begin{table}
\centering
\def\~{\hphantom{0}}
\caption{Summary of functional model-performance measure combinations used in multiple linear regression models.}
\label{tab:multlinreg}
\begin{tabular*}{\columnwidth}{@{}c@{\extracolsep{\fill}}l@{\extracolsep{\fill}}}
\Hline
{\textbf{Combination}} & {\textbf{Covariates}}\\
\hline
1 & LCSG; D; E \\
2 & LCSG; D; EN \\
3 & LCSG; D; EN; CL \\
4 & CL; PCL; PNS \\
5 & LCSG; D; EN; CL; PCL; PNS \\
6 & LCSG; D; EN; CL; PCL; PNS; DC \\
\Hline
\end{tabular*}
\end{table}


We then test the predictive accuracy of each of the 90 resulting regression models using repeated random holdout validation.  For each model, we randomly split our initial data into two sets: training data (90\% of initial data) and validation data (10\% of initial data). We use our training data to fit a regression model using the initial combination of parameters from Tables~\ref{tab:simplinreg} and \ref{tab:multlinreg}. We then use this new regression model to predict load curtailed for each record in the validation data set.  We compare these predicted values to the actual values from our AC load flow analysis. For each of the 90 full regression models, we repeat this process 100 times (beginning with the random split of our initial data) for a 100-fold random holdout cross-validation.


\section{Results}
\label{sec:results}

\begin{figure*}
\centerline{\includegraphics[width=\textwidth]{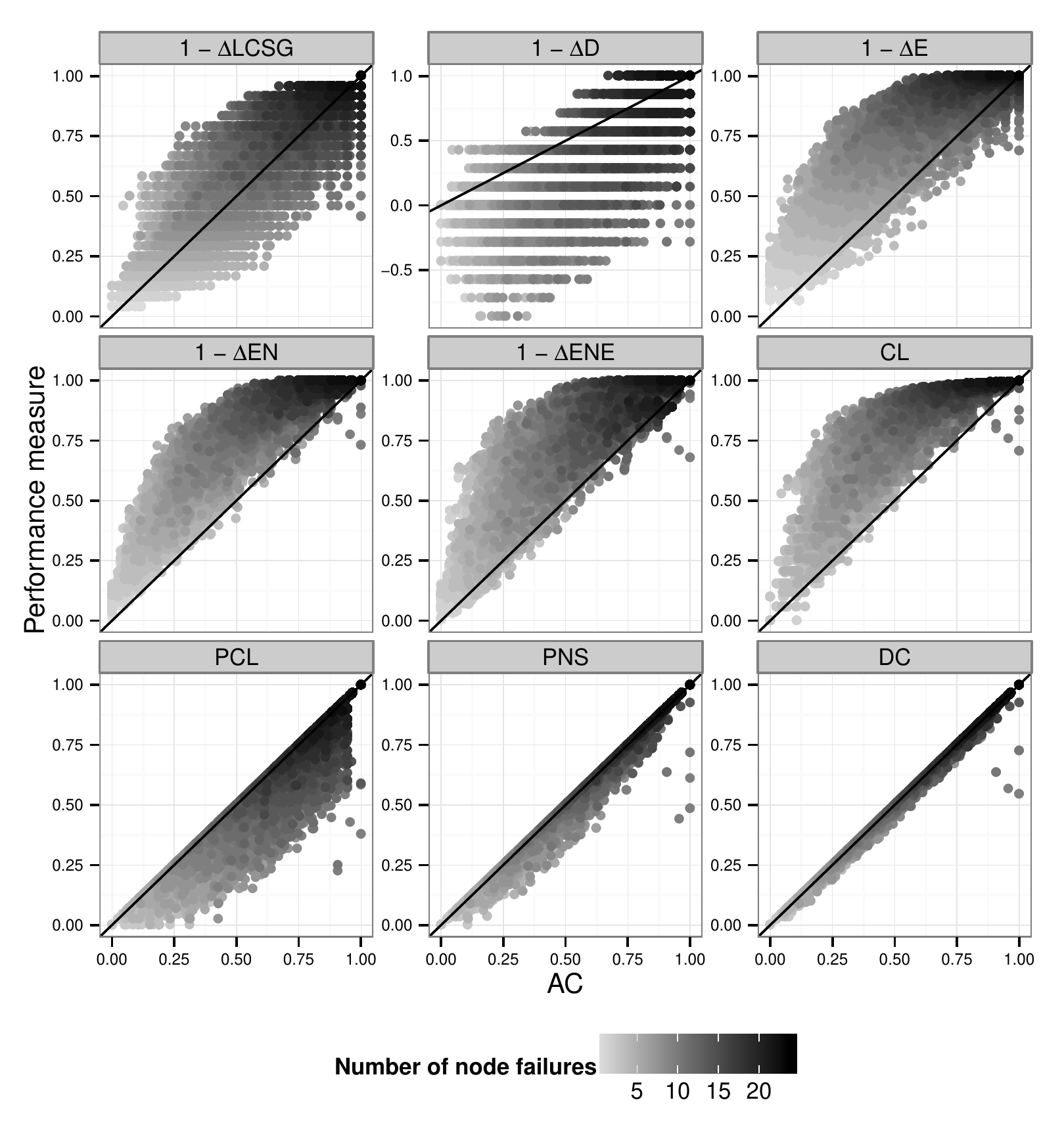}}
\caption{Correlation plots for node removals. Each dot represents the system performance for a given failure scenario as calculated by a given functional model-performance measure (y-axis) and the AC load flow analysis (x-axis).}
\label{fig:pfScatterNodes}
\end{figure*}

\begin{figure*}
\centerline{\includegraphics[width=\textwidth]{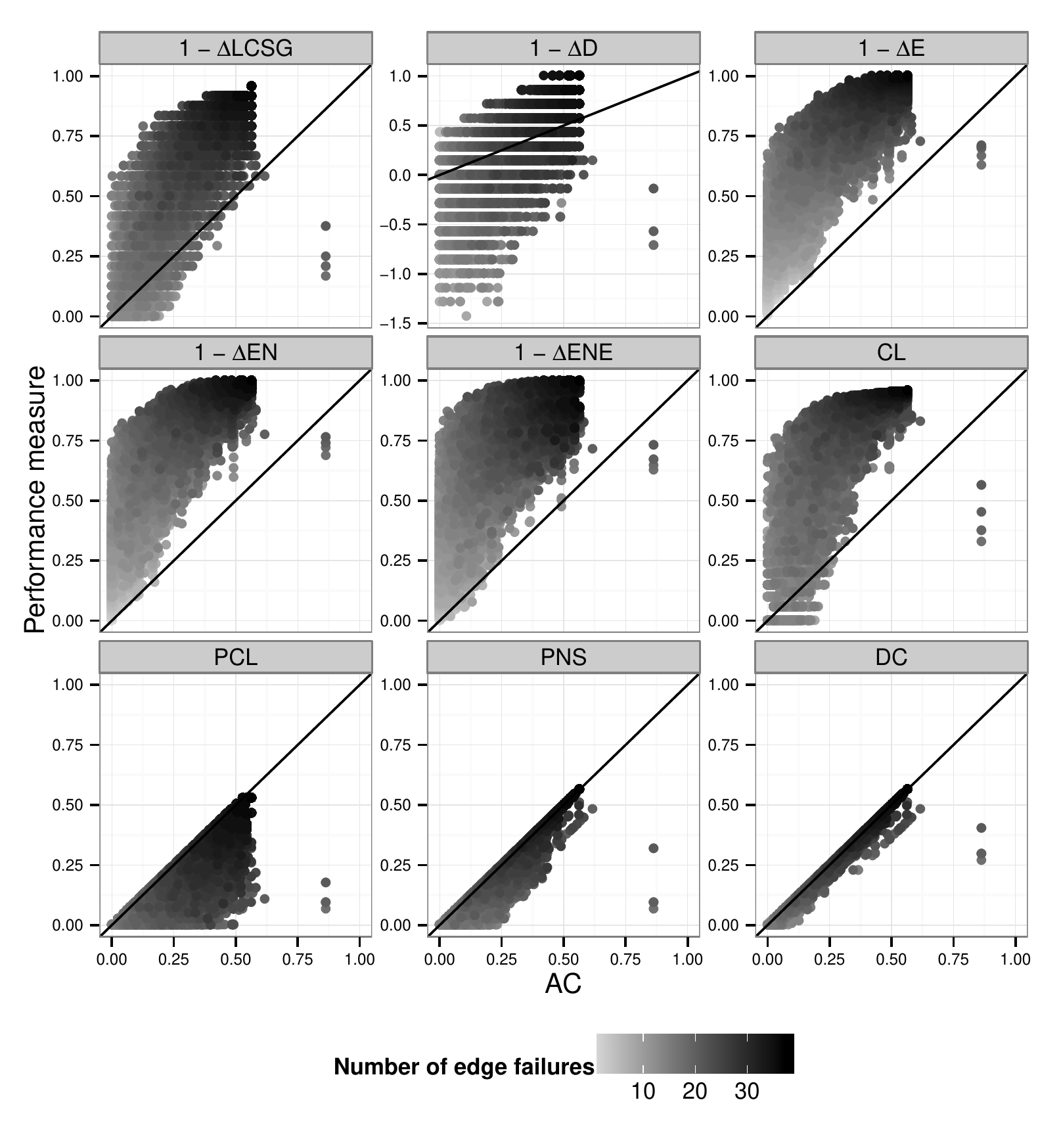}}
\caption{Correlation plots for edge removals. Each dot represents the system performance for a given failure scenario as calculated by a given functional model-performance measure (y-axis) and the AC load flow analysis (x-axis).}
\label{fig:pfScatterEdges}
\end{figure*}

For each node failure scenario and edge failure scenario, we use our functional model-performance measures to assess the behavior of the system after each level of component removal (i.e., 1 component removed, 2 components removed, through $n$ components removed, where $n$ is the number of nodes or edges in the system).  Figures~\ref{fig:pfScatterNodes} and~\ref{fig:pfScatterEdges} present comparisons of each functional model-performance measure with the results of the AC load flow analysis for all failure scenarios and numbers of components removed. Based on these results, it is clear that although a functional model-performance measure may give a reasonable estimate of the average consequences that arise, the correctness of the estimate for the individual scenarios may vary greatly. This is significant, because in reality systems are not typically subjected repeatedly to varying failure scenarios. Instead, when assessing system robustness, it is important to understand reasonably well how the system will 
perform when subjected to a specific failure scenario, and unfortunately this information is not provided by all functional model-performance measures.  

Figures~\ref{fig:pfScatterNodes} and~\ref{fig:pfScatterEdges} show that the accuracy of the performance measures largely follows the classification in Section~\ref{sec:models}; that is, in general, the greater the inclusion of functional characteristics, the better the estimate of the system's actual performance for a given failure scenario. The topological performance measures LCSG and D both significantly overestimate and underestimate the consequences for individual failure scenarios, though the diameter measure more often underestimates consequences. One reason that the largest connected subgraph measure may overestimate consequences is that it is possible for the system to split into two subgraphs, or islands, but still be able to supply all the load from the generators in each island. In such a situation, the LCSG performance measure would estimate significantly decreased performance, when in fact the system was still functioning at its initial performance. 

The performance measures E, EN, ENE, and CL typically overestimate consequences as compared to the AC model. The more physically oriented models, PCL, PNS and DC nearly always underestimate the consequences for individual scenarios as compared to the AC model. The reason for this is because they do not account for voltage and branch constraints (except for the DC optimal power flow model, which does consider active power flow branch constraints). The proposed performance measures EN and ENE do not capture the behavior of the system better than the classic network theoretic measure of efficiency (E), which does not take any physical aspects into account. The best performing functional model-performance measures are clearly PCL, PNS and DC, but LCSG appears to also give a reasonable estimate of system performance for node removals, but less so for edge removals.

\begin{sidewaysfigure}
\centerline{\includegraphics[width=600pt]{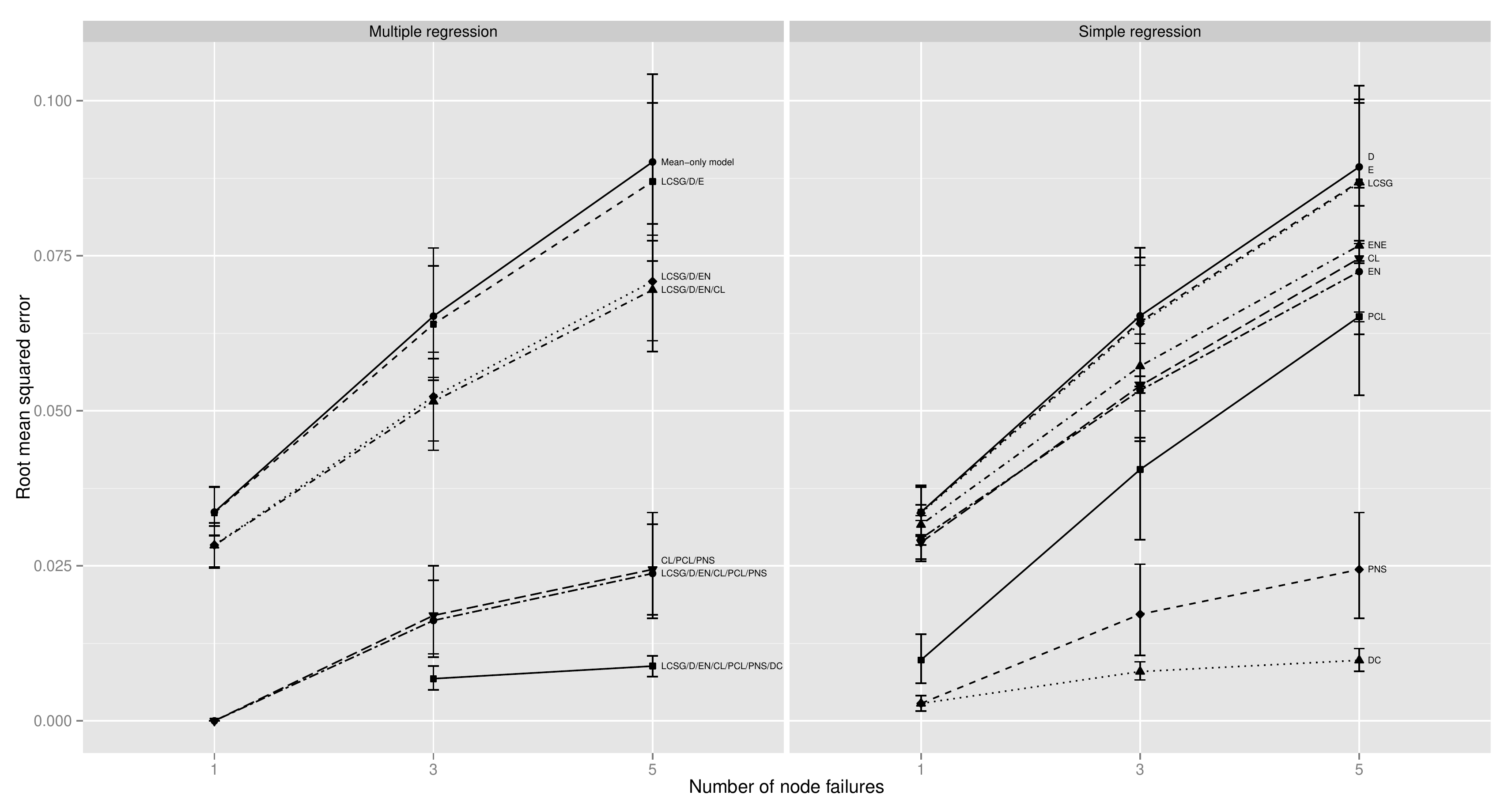}}
\caption{Root mean squared errors for predictions of system performance after node failures based on 100 holdout samples using simple and multiple regression models. Error bars give 95\% confidence intervals.}
\label{fig:pfHoldoutNodes}
\end{sidewaysfigure}

\begin{sidewaysfigure}
\centerline{\includegraphics[width=600pt]{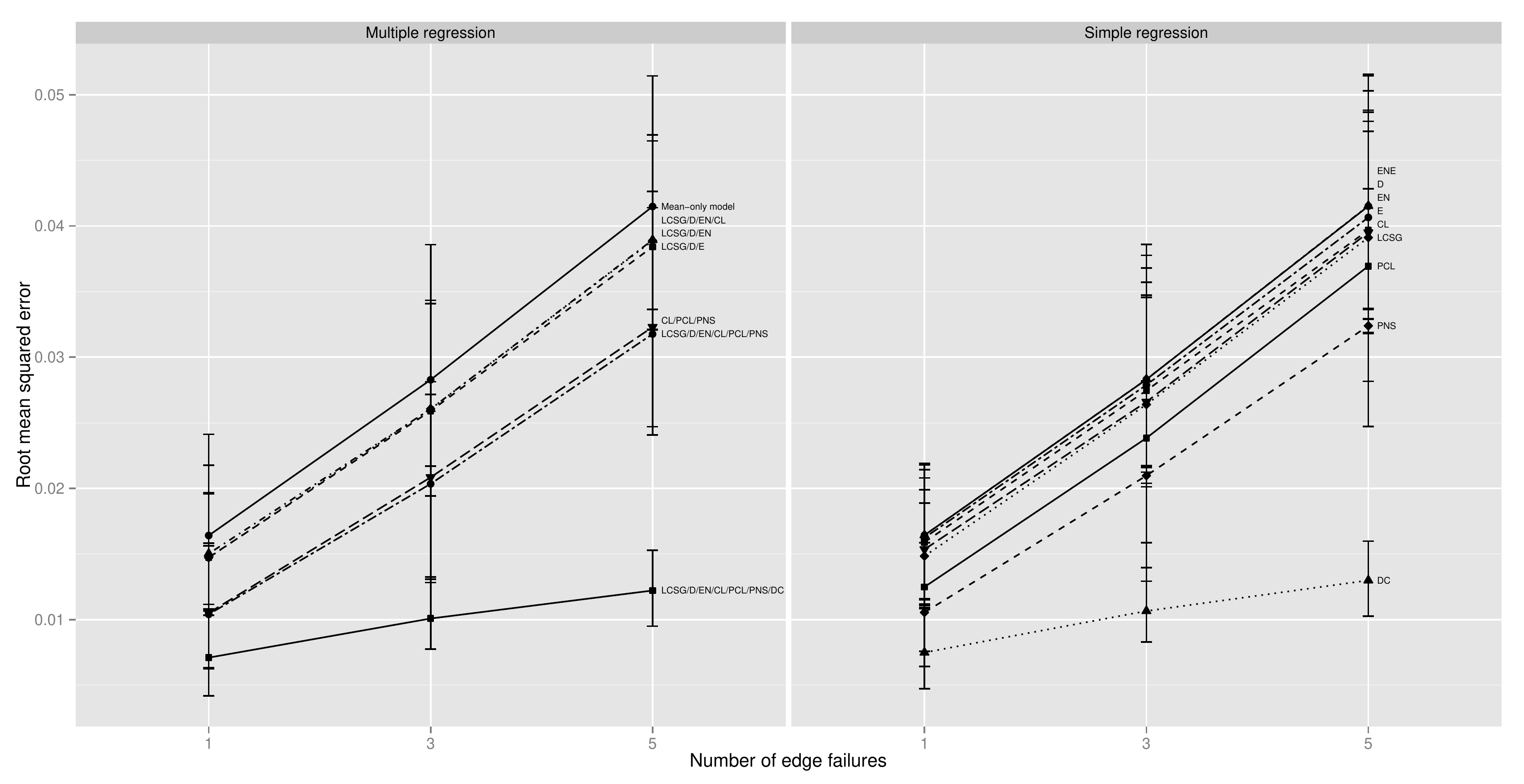}}
\caption{Root mean squared errors for predictions of system performance after edge failures based on 100 holdout samples using simple and multiple regression models. Error bars give 95\% confidence intervals.}
\label{fig:pfHoldoutEdges}
\end{sidewaysfigure}

The repeated random holdout validation tests conducted for each of our 90 regression models (45 models for node failures and 45 models for edge failures) further support the trends described above.  That is, we see that when more physical information about the system that is included in a single or group of functional model-performance measure(s), these functional model-performance measure(s) are, in general, better able to predict AC-load curtailed. Figures~\ref{fig:pfHoldoutNodes} and \ref{fig:pfHoldoutEdges} present the root mean squared errors averaged over 100 holdout samples for each of the regression models; the error bars represent 95\% confidence intervals.

For node failure scenarios, we see that the three topological models with undifferentiated components (D, E, and LCSG) result in the highest predictive errors.  The topological models with differentiated components (ENE, CL, and EN) provide slightly better estimates of AC-load curtailed.  The simplistic capacity models (PCL and PNS) and the physical flow model (DC) have significantly lower predictive errors than either category of topological models.  Of particular interest here is the relatively high predictive accuracy of the simplistic capacity model, Power Not Supplied (PNS).  This functional model does not require complete modeling of physical flows, yet it is still able to estimate the AC behavior of the system significantly better than the simpler topological models.  However, it is important to note that even the most complicated functional model, the DC load flow model, has a non-zero predictive error and is not able to completely capture the true behavior of the system.

Similar patterns appear when using multiple functional model-performance measures to predict system behavior.  Combinations of functional model-performance measures encompassing less physical information about the system have lower predictive accuracy (\emph{i.e.,} higher predictive error) than combinations that include more physical details. Several combinations of functional model-performance measures are particularly interesting here.  The LCSG/D/EN/CL regression model uses only topologically-based functional models, so it is fairly simple both with respect to computation and data requirements. However, combining these functional model-performance measures provides an increase in predictive accuracy over any of the single functional model-performance measures; this increase becomes larger as the number of node failures increases.  The LCSG/D/EN/CL/PCL/PNS regression model combines topological models with simplistic capacity models; this combination of functional models-performance measures also increases 
the predictive accuracy of the regression model over any of the single function model-performance measures.  The increase is particularly significant when only one node fails in a given scenario, bringing the predictive error of the model close to zero.

Results for the edge failure scenarios are similar to those for the node failure scenarios.  The topologically-based functional models again have high predictive error, but here there is less distinction between the predictive accuracy of topological models with differentiated and undifferentiated components.  The simplistic capacity models (PCL and PNS) have lower predictive error than the topological functional models, though PNS does not provide as large an improvement over PCL for edge failures as it did for node failures.  This difference may arise in part because simplistic capacity models do not incorporate capacity constraints for power lines; such constraints are likely to have a more significant effect on system behavior when it is subjected to edge failures than when it experiences node failures as fewer lines are available in the system to carry the power from generators to load.  Finally, as with the node failure scenarios, the DC load flow functional model results in significant predictive 
error.

Overall, when combining multiple functional model-performance measures to predict system behavior for edge failures, we see larger improvements over single functional model-performance measure predictions than we see with node failures.  Here, all three combinations of topological functional model-performance measures (LCSG/D/E; LCSG/D/EN; LCSG/D/EN/CL) provide higher predictive accuracy for each level of edge removal than do any of the included single topological functional model-performance measures.  Again, because these functional model-performance measures are computationally simple, the benefits in increased predictive accuracy gained by combining several functional model-performance measures do not come at a high cost. Combining topological and simplistic capacity functional model-performance measures (CL/PCL/PNS; LCSG/D/EN/CL/PCL/PNS) also results in higher predictive accuracy than any of the included functional model-performance measures individually.


\section{Discussion}
\label{sec:discussion}

The results here clearly depict that the greater the inclusion of physical characteristics in the functional model, the better the estimate of the system’s actual performance when perturbed. Using more complicated performance measures does come at a cost, primarily in computational time but also with regards to the information about the system that is required. In our analysis, mean simulation times for a given node failure scenario ranged from 0.1 (0.1 for edges) seconds \footnote{Simulations were performed using a single core of an Intel Xeon 5160 quad core 3.00 GHz processor.} for the simplest topological approaches to 1.2 (3.8) seconds for the DC load flow model and 3.5 (10.8) seconds for the AC load flow model. At first glance, these simulation times may all seem quite reasonable, but it is important to note that our test system is much smaller than real-world systems, and differences in simulation times between simple and advanced approaches will scale exponentially.

The results shown in this paper do not imply that the more simplistic performance measures do not provide any useful information. As has been shown, several topologically-based performance measures that also include some physical information (i.e., power connection loss (PCL) and power not supplied (PNS)) provide similar results to the DC and AC load flow models in some situations. These measures are likely to provide reasonable representations of reality in complex, large-scale modeling situations in which physical flow modeling is prohibitively time-consuming.

The results in this paper are based on a single test power system that is quite small in size. Therefore, in the future it may be beneficial to perform similar studies of power systems with a much larger number of components, such as the IEEE 300 bus system or the Western Interconnection of the United States. This would aid in validating the general conclusions drawn in the present paper, but would also provide insight as to how the simulation times for the different performance measures scale with the size of the system. Furthermore, it would be of interest to compare power systems of different types (e.g. transmission, sub-transmission, and distribution) to see how the performance measures described in this paper behave for these. Finally, this research can be extended to include similar studies for other types of critical infrastructures, such as water supply systems, telecommunication systems, and transport systems.


\section{Conclusions}
\label{sec:conclusions}

This paper presents a classification for different types of functional models that can be used for risk and vulnerability analysis of electric power systems. These approaches span from very simple topologically-oriented models to advanced models based on the engineering and physics of flows in the system. In order to compare the performance estimates achieved by these different types of functional models and performance measures, we performed a simulation study using the IEEE RTS 96 test power system. From our study, we conclude that while some performance measures may capture the average behavior of the system when perturbed, the accuracy of the performance estimates for specific scenarios may vary greatly. In other words, topology-based measures are of limited value in analyzing the robustness of particular power systems under specific failure scenarios. Hence, great care should be taken when using these types of approaches as inputs to decision-making for managing power system vulnerabilities. On the 
other hand, simplistic approaches sometimes allow for analysis of a broad spectrum of scenarios when assessing system vulnerability when such a range of scenarios may be too difficult to model with more complex methods. Accurate models of infrastructure performance are critical for infrastructure risk and vulnerability analysis, and further studies are needed to understand the trade-offs between fidelity and complexity for performance models for other types of critical infrastructure systems such as water, communication, and transportation systems. 


\backmatter

\section*{Acknowledgments}





\begin{thebibliography}{99}

\bibitem{Rinaldi2004}
Rinaldi SM. Modeling and simulating critical infrastructures and their interdependencies. In: Proceedings of the 37th Hawaii International Conference on System Sciences; 2004; Hawaii.

\bibitem{Watts1998}
Watts DJ, Strogatz SH. Collective dynamics of `small-world' networks. Nature, 1998; 393:440-442.

\bibitem{Motter2002b}
Motter A, Nishikawa T, Lai Y-C. Range-based attack on links in scale-free networks: Are long-range links responsible for the small-world phenomenon?. Physical Review E, 2002; 66(6):065103.

\bibitem{Crucitti2004a}
Crucitti P, Latora V, Marchiori M. A topological analysis of the Italian electric power grid. Physica A: Statistical Mechanics and its Applications, 2004; 338:92-97.

\bibitem{Chassin2005}
Chassin DP, Posse C. Evaluating North American electric grid reliability using the Barab\'{a}si-Albert network model. Physica A: Statistical Mechanics and its Applications, 2005; 355:667-677.

\bibitem{Holmgren2006a}
Holmgren \AA. Using graph models to analyze the vulnerability of electric power networks. Risk Analysis, 2006; 26(4):955-969.

\bibitem{Wang2009a}
Wang J-W, Rong L-L. Cascade-based attack vulnerability on the US power grid. Safety Science, 2009; 47(10):1332-1336.

\bibitem{Johansson2012}
Johansson J, Hassel H. Modelling, simulation and vulnerability analysis of interdependent technical infrastructures. In: Hokstad P, Utne IB, Vatn J, editors. Risk and interdependencies in critical infrastructures: A guideline for analysis. London: Springer-Verlag; 2012. p. 49-66.

\bibitem{Grubesic2008}
Grubesic TH, Matisziw TC, Murray AT, Snediker D. Comparative approaches for assessing network vulnerability. International Regional Science Review, 2008; 31(1):88-112.

\bibitem{Hines2010}
Hines P, Cotilla-Sanchez E, Blumsack S. Do topological models provide good information about electricity infrastructure vulnerability?. Chaos, 2010; 20(3):033122.

\bibitem{Overbye2004}
Overbye TJ, Cheng X, Sun, Y. A comparison of the AC and DC power flow models for LMP calculations. In: Proceedings of the 37th Hawaii International Conference on System Sciences; 2004; Hawaii.

\bibitem{Chen2010}
Chen G, Dong ZY, Hill DJ, Zhang GH, Hua KQ. Attack structural vulnerability of power grids: A hybrid approach based on complex networks. Physica A: Statistical Mechanics and its Applications, 2010; 389:595-603.

\bibitem{Holme2002}
Holme P, Kim B, Yoon C, Han S. Attack vulnerability of complex networks. Physical Review E, 2002; 65(5):056109.

\bibitem{Latora2005}
Latora V, Marchiori M. Vulnerability and protection of infrastructure networks. Physical Review E, 2005; 71(1):015103.

\bibitem{Rosas-Casals2007}
Rosas-Casals M, Valverde S, Sol\`{e} RV. Topological vulnerability of the European power grid under errors and attacks. International Journal of Bifurcation and Chaos, 2007; 17(7):2465-2475.

\bibitem{Winkler2010}
Winkler J, Due\~{n}as-Osorio L, Stein R, Subramanian D. Performance assessment of topologically diverse power systems subjected to hurricane events. Reliability Engineering and System Safety, 2010; 95(4):323-336.

\bibitem{Albert2004}
Albert R, Albert I, Nakarado G. Structural vulnerability of the North American power grid. Physical Review E, 2004; 69(2):025103.

\bibitem{Duenas-Osorio2009}
Due\~{n}as-Osorio L, Vemuru SM. Cascading failures in complex infrastructure systems. Structural Safety, 2009; 31(2):157-167.

\bibitem{Wang2011}
Wang K, Zhang B-H, Zhang Z, Yin X-G, Wang B. An electrical betweenness approach for vulnerability assessment of power grids considering the capacity of generators and load. Physica A: Statistical Mechanics and its Applications, 2011; 390:4692-4701.

\bibitem{Jonsson2008}
J\"{o}nsson H, Johansson J, Johansson H. Identifying critical components in technical infrastructure networks. Journal of Risk and Reliability, 2008; 222(2):235-243.

\bibitem{Dobson2001}
Dobson I, Carreras BA, Lynch VE, Newman DE. An initial model for complex dynamics in electric power system blackouts. In: Proceedings of the 34th Hawaii International Conference on System Sciences; 2001; Maui, Hawaii.

\bibitem{Carreras2002a}
Carreras, BA, Lynch VE, Dobson I, Newman DE. Critical points and transitions in an electric power transmission model for cascading failure blackouts. Chaos, 2002; 12(4):985-994.

\bibitem{Dobson2002}
Dobson I, Chen J, Thorp JS, Carreras BA, Newman DE. Examining criticality of blackouts in power system models with cascading events. In: Proceedings of the 35th Hawaii International Conference on System Sciences; 2002; Hawaii.

\bibitem{Dobson2003}
Dobson I, Carreras BA, Newman DE. A probabilistic loading-dependent model of cascading failure and possible implications for blackouts. In: Proceedings of the 36th Hawaii International Conference on System Sciences; 2003; Hawaii.

\bibitem{Song2005}
Song H, Kezunovic M. Static security analysis based on vulnerability index (VI) and network contribution factor (NCF) method. In: Proceedings of the Transmission and Distribution Conference and Exhibition: Asia and Pacific; 2005.

\bibitem{Pepyne2007}
Pepyne DL. Topology and cascading line outages in power grids. Journal of Systems Science and Systems Engineering, 2007; 16(2):202-221.

\bibitem{Arianos2009}
Arianos S, Bompard E, Carbone A, Xue F. Power grid vulnerability: a complex network approach. Chaos, 2009; 19(1):013119.

\bibitem{Grigg1999}
Grigg C, Wong P, Albrecht P, Allan R, Bhavaraju M, Billinton R, Chen Q, Fong C, Haddad S, Kuruganty S, Li W, Mukerji R, Patton D, Rau N, Reppen D, Schneider A, Shahidehpour M, Singh C. The IEEE Reliability Test System - 1996. IEEE Transactions on Power Systems, 1999; 14(3):1010-1020.

\bibitem{Sole2008}
Sol\'{e} RV, Rosas-Casals M, Corominas-Murtra B, Valverde S. Robustness of the European power grids under intentional attack. Physical Review E, 2008; 77(2):026102.

\bibitem{Hines2011}
Hines P, Cotilla-Sanchez E, Blumsack S. Topological models and critical slowing down: Two approaches to power system blackout risk analysis. In: Proceedings of the 44th Hawaii International Conference on System Sciences; 2011; Hawaii.

\bibitem{Johansson2007}
Johansson J, J\"{o}nsson H, Johansson H. Analysing the vulnerability of electric distribution systems: A step towards incorporating the societal consequences of disruptions. International Journal of Emergency Management, 2007; 4(1):4-17.

\bibitem{Zimmerman2011}
Zimmerman RD, Murillo-S\'{a}nchez CE, Thomas RJ. MATPOWER: steady-state operations, systems research and education. IEEE Transactions on Power Systems, 2011; 26(1):12-19.

\bibitem{IEEE1995}
Price WW, Taylor CW, Rogers GJ, Srinivasan K, Concordia C, Pal MK, Bess KC, Kundur P, Agrawal BL, Luini JK, Vaahedi E, Johnson BK. Standard load models for power flow and dynamic performance simulation. IEEE Transactions on Power Systems, 1995; 10(3):1302-1313.

\bibitem{Billinton1996}
Billinton R, Allan RN. Reliability evaluation of power systems. 2nd ed. New York: Springer; 1996.

\bibitem{Arnborg1997}
Arnborg S, Andersson G, Hill DJ, Hiskens IA. On undervoltage load shedding in power systems. Electrical Power \& Energy Systems, 1997; 19(2):141-149.

\bibitem{Arini1999}
El Arini, M. Optimal dynamic load shedding policy for generation load imbalances including characteristics of loads. International Journal of Energy Research, 1999; 23:79-89.

\bibitem{Ladhani2004}
Ladhani SS, Rosehart W. Under voltage load shedding for voltage stability overview of concepts and principles. In: Proceedings of the Power Engineering Society General Meeting; 10 June 2004; Denver, Colorado. IEEE; p. 1597-1602.

\bibitem{Taylor1994}
Taylor, CW. Power system voltage stability.  McGraw-Hill Companies; 1994.

\end{thebibliography}
\end{document}